\documentclass[sigconf, hidelinks]{acmart}
\usepackage{graphicx}

\usepackage{amsmath}
\usepackage{comment}
\usepackage{tabularx}
\usepackage{multirow, multicol}
\newcolumntype{Y}{>{\centering\arraybackslash}X} 
\usepackage{tabularray}
\usepackage{makecell}
\usepackage{array}
\usepackage{xspace}
\usepackage[font=small]{caption}
\usepackage{orcidlink}
\usepackage{nicefrac}
\usepackage{siunitx}
\usepackage{float}
\usepackage{dblfloatfix}
\usepackage{enumitem}
\usepackage{subcaption}
\usepackage{relsize}
\usepackage{adjustbox}
\usepackage{soul}
\usepackage{amsthm}
\usepackage{mathtools}
\usepackage{bbm}
\usepackage{etoolbox}
\usepackage{titlesec}
\usepackage{diagbox}
\usepackage[hang, flushmargin]{footmisc}

\makeatletter \def\ClassError#1#2#3{\ClassWarning{#1}{#2}} \makeatother

\def\sothat{\,\vert\,}
\newcommand\abs[1]{\left\lvert#1\right\rvert}\newcommand{\concat}{\ensuremath{\,+\!\!+\,}}

\newcommand{\kk}{\text{\scriptsize\sf k}\xspace}
\newcommand{\MM}{\text{\scriptsize\sf M}\xspace}

\DeclareMathOperator*{\argmax}{argmax}
\DeclareMathOperator*{\argmin}{argmin}

\def\hedge{\mbox{h\!\!\;-\!\!\;edge}\xspace}
\def\hgraph{\mbox{h\!\!\;-\!\!\;graph}\xspace}
\def\hedges{\mbox{h\!\!\;-\!\!\;edges}\xspace}
\def\hgraphs{\mbox{h\!\!\;-\!\!\;graphs}\xspace}

\def\Hgraph{\mbox{H\!\!\;-\!\!\;graph}\xspace}

\newskip\paragraphbreak
\DeclareRobustCommand{\semibold}{\fontseries{sb}\selectfont\SetTracking{encoding=*}{-54}\lsstyle}
\DeclareTextFontCommand{\textsemibold}{\semibold}

\titleformat{\subsubsection}[runin]
  {\itshape\semibold}
  {\thesubsubsection}{1em}{}[.]

\AtBeginDocument{\AddToHook{cmd/section/before}{\vspace{-3pt plus 3pt minus 0pt}}}
\AtBeginDocument{\AddToHook{cmd/subsection/before}{\vspace{-4pt plus 4pt minus 0pt}}}
\AtBeginDocument{\AddToHook{cmd/subsubsection/before}{\vspace{-4pt plus 4pt minus 0pt}}}

\expandafter\def\expandafter\normalsize\expandafter{%
    \normalsize
    \setlength\abovedisplayskip{3.5pt} 
    \setlength\belowdisplayskip{3.5pt} 
    \setlength\abovedisplayshortskip{2pt} 
    \setlength\belowdisplayshortskip{2pt} 
}

\newlength{\itemizepadding}
\setlist[itemize]{topsep=\itemizepadding, leftmargin=2.5em} 
\setlist[enumerate]{topsep=\itemizepadding, leftmargin=2.5em} 

\begin{document}

\copyrightyear{2027}
\acmYear{2027}
\setcopyright{none}
\acmConference[N/A]{N/A}{N/A}{N/A}
\acmBooktitle{N/A}
\acmDOI{}
\acmISBN{}

\renewcommand\footnotetextcopyrightpermission[1]{}

\title{GPU-Accelerated Hypergraph Partitioning and Placement to~Map~SNNs on Neuromorphic Hardware}

\author{Marco Ronzani}
\orcid{0009-0002-8485-0717}
\affiliation{%
  \institution{DEIB, Politecnico di Milano}
  \city{Milan \vspace{-2pt}}
  \country{Italy \vspace{-2pt}}
}

\author{Cristina Silvano}
\orcid{0000-0003-1668-0883}
\affiliation{%
  \institution{DEIB, Politecnico di Milano}
  \city{Milan \vspace{-2pt}}
  \country{Italy \vspace{-2pt}}
}

\renewcommand{\shortauthors}{Anonymous}

\begin{abstract}
    SNNs running on neuromorphic hardware use spikes to achieve sparse and energy-efficient communication over a mesh of cores.
    In turn, system performance heavily depends on the assignment of neurons to cores: the mapping.
    Since hardware features inter-core multicast and intra-core replication of spikes, we model SNNs as hypergraphs to exploit both opportunities for reducing communication traffic.
    Mapping thus comprises two NP-hard problems: hypergraph partitioning and placement on the lattice of cores.
    High-quality solutions to both are critical, yet increasingly difficult as networks scale to millions of neurons.
    Therefore, we propose a GPU-accelerated pipeline for SNN mapping: a multi-level partitioning scheme is devised around hardware constraints, while placement is initialized through recursive bisection, followed by refinement pulling together strongly connected cores through repeated swaps.
    Model-based experiments show upwards of $16\%$ lower latency and $42\%$ lower energy for spike movements over existing sequential tools, while our parallel mapper is on average $18$-$280\times$ faster.%
\end{abstract}

\maketitle

\vspace{-6pt}

\section{Introduction}\label{sec:intro}

Spiking Neural Networks (SNNs) are a brain-inspired, event-based artificial intelligence paradigm with major potential for energy efficiency and temporal awareness \cite{SNNSurvey, SNNvsANN}.
These advantages are at their best when SNNs are run on NeuroMorphic Hardware (NMH), dedicated accelerators designed to leverage their sparse activity \cite{TrueNorthDesign, SNNHardwareImplementations, NeuromorphicHardwareSurvey}.
Recently, SNNs exceeded ten million neurons and are heading toward brain-scale \cite{MouseOnFugaku, NatureBrainScaleSNNonHPC}, while NMH accelerators already manage several million neurons \cite{NeuromorphicComputingAtScale}.
Between network and hardware, however, lies the challenge of determining how the former should execute on the latter: the \textbf{mapping problem}.
Its complexity grows with system size, and final runtime performance depends on the quality of its solution \cite{DFSynthesizer, MappingVeryLargeSNNtoNHW}.

A SNN comprises neurons whose axons connect to downstream neurons through multiple synapses; information propagates as spikes over these one-to-many connections.
Accordingly, we model SNNs as \textbf{hypergraphs}, one axon, one hyperedge \cite{AxonFlow}.
The performance for running a SNN on NMH directly depends on the distance traveled by spikes between cores, as every hop adds energy and time to its delivery \cite{TrueNorthEcosystem, MappingVeryLargeSNNtoNHW}.
Hence, the role of a mapping is to assign neurons to cores while minimizing the ensuing spike traffic.

Formally, mapping reduces to two well-known problems: hypergraph \textbf{partitioning}, packing neurons into core-compliant partitions, and hypergraph \textbf{placement}, assigning partitions to cores.
The two minimize the volume of spikes moving around and the path length they cover, respectively.
Both problems are notoriously NP-hard and must be solved via heuristics \cite{AdvancesInHypergraphPartitioning, KaHIPPlacement}.

Several approaches exist to handle these problems, both in general and specifically for SNN mapping.
Classic works on hypergraph partitioning are hMETIS~\cite{hMETIS_vlsi}, KaHyPar~\cite{KaHyPar, MtKaHyPar}, and BiPart~\cite{BiPart}.
Placement has been widely studied for process assignment in parallel computing \cite{MPIPlacement, KaHIPPlacement}, and VLSI \cite{WirelengthHypergraphPlacement, ForestAndSteinerTrees}.
Yet all such methods lack support for the costs and constraints of NMH, instead focusing on $k$-way balanced partitioning and quadratic assignment problem variants.
Therefore, specialized SNN mapping algorithms have been proposed, like DFSynthesizer \cite{DFSynthesizer}, EdgeMap \cite{EdgeMap}, SNNcut \cite{SNNcut, MappingVeryLargeSNNtoNHW, MappingVeryLargeSNNtoNHWv2}, and others \cite{SpiNeMap, NSGASnnMapping, PACMAN, SNEAP}.
However, these tools had to forgo the former's optimized heuristics in favor of greedier, faster alternatives.
Otherwise, scaling to networks with millions of neurons would lead to days of execution time \cite{AxonFlow}.
Yet, to date, their algorithms remain inherently sequential and implemented on CPU.

As such, massively parallel partitioning and placement heuristics become a compelling approach when scaling SNNs and NMH.
A few exist, namely HyperG~\cite{HyperG}, gHyPart~\cite{gHyPart}, and Frishman et al. \cite{GraphPlacementGPU}.
Though, once more, they target different problem formulations, precluding their use in SNN mapping.
Nonetheless, they show attractive results, exceeding a $100\times$ speedup over \mbox{sequential execution}.

Following this direction, we develop GPU algorithms for hypergraph partitioning and placement tailored to SNN mapping on NMH.
Massive parallelism is used to efficiently handle the problem at its largest, during partitioning, and then simultaneously explore many placements.
This enables mapping SNNs with hundreds of millions of synapses in minutes without sacrificing solution quality.


Altogether, to the best of our knowledge, this work presents the first GPU-accelerated pipeline for mapping spiking neural networks on neuromorphic hardware.
In particular, it introduces:
\begin{enumerate}
    \item a multi-level hypergraph partitioning algorithm specialized for NMH constraints and minimizing spike traffic volume.
    \item a multi-start placement routine using recursive bisection to produce a high-locality assignment of neuron groups, followed by force-directed refinement of their hardware layout.
\end{enumerate}
Tested on 12 SNNs ranging from 0.8\MM to 577\MM synapses, our algorithms achieve mappings with mean analytical costs of $0.58\times$ energy consumption and $0.84\times$ average spike delivery latency compared to the best from SoTA tools.
Meanwhile, on a per-tool average, our mapping construction is between $18$-$280\times$ faster.

\section{Preliminaries}\label{sec:preliminaries}

\vspace{2pt}

\subsubsection*{Spiking Neural Networks}\label{subsec:snns}

A SNN is composed of neurons connected via axons and synapses.
Neurons integrate incoming spikes and, upon crossing a threshold, emit a spike along their axon to all downstream neurons.
The network, per se, is asynchronous, though it is simulated in discrete spike propagation steps.
Consequently, information is encoded via the timing or rate of spike events \cite{SNNSurvey, SNNHardwareImplementations}.

Several kinds of SNNs exist.
Some are derived from Artificial Neural Networks (ANNs), achieving similar accuracy in static domains.
Others are trained natively, a topic of ongoing research with promising results on event-driven tasks \cite{SNNvsANN}.
These two families differ a lot in topology.
ANN-based networks are feedforward, thus acyclic and with local, regular connections.
Instead, native ones own a small-world structure, with cycles and dense neighborhoods~\cite{AxonFlow}.

\subsubsection*{Neuromorphic Hardware}\label{subsec:neuromorphic_hw}

NMH mimics the distributed, event-driven nature of SNNs to accelerate their simulation.
It comprises a mesh of cores, each hosting several neurons and connected over a network-on-chip that circulates spikes.
Typical core arrangements are a 2D lattice or a toroid \cite{Neurogrid, SpiNNaker}.
Crucial to its power efficiency, most of the hardware is active solely upon receiving a spike.
Simulation speed is instead bound by spike delivery latency.
Hence, performance depends on communication, being governed by the volume of transmitted spikes and their travel distance \cite{TrueNorthDesign, NeuromorphicHardwareSurvey}.

Central to NMH are two mechanisms that exploit the one-to-many propagation of spikes to reduce communication traffic: intra-core replication and inter-core multicast.
When two neurons slated to receive the same spike are in the same core, the spike is delivered to the core only once, and replicated internally.
When two destination neurons are in different cores, instead, spike multicast is used over the interconnect to progressively fan it out.

Notable instances of NMH are TrueNorth~\cite{TrueNorthDesign}, Loihi~\cite{Loihi}, SpiNNaker~\cite{SpiNNaker}, and Neurogrid \cite{Neurogrid}, but many others exist \cite{Darwin3, Tianji, SENECA, NeuromorphicHardwareSurvey}.

\subsubsection*{SNN and Hardware Models}\label{subsec:network_and_hardware_modeling}

\begin{figure}[t]
    \centering
    \vspace{-0pt} 
    \includegraphics[width=1.0\columnwidth]{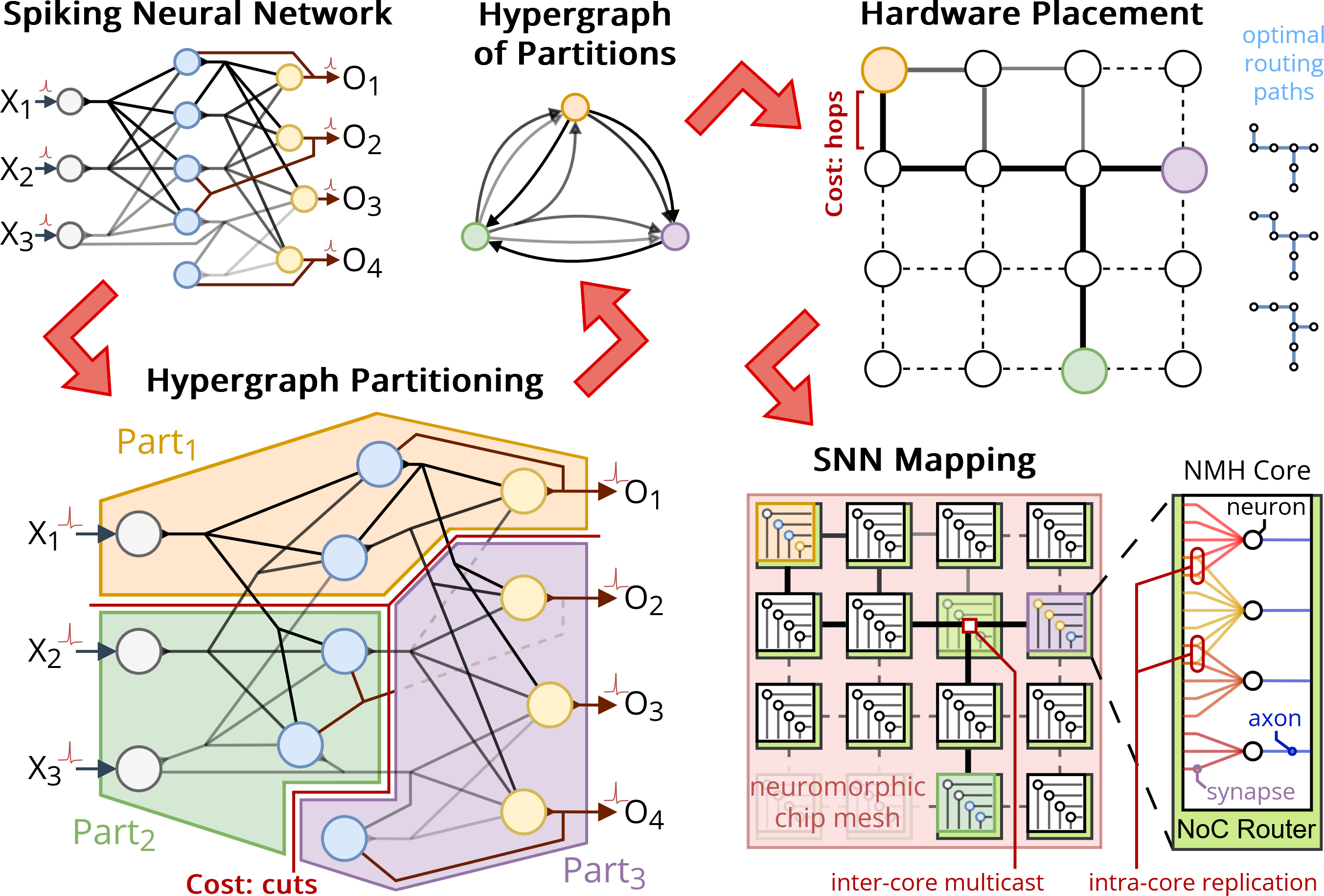} 
    \vspace{-18pt}
    \caption{SNN mapping problem overview.}
    \vspace{-3pt}
    \label{fig:mapping_overview}
\end{figure}

By representing each neuron as a node and its axon as a hyperedge (\hedge) with a pin per synapse, SNNs are naturally modeled by a hypergraph (\hgraph).
A hypergraph generalizes the concept of graph with edges --~\hedges~-- connecting more than two nodes.
In particular, a SNN is a weighted directed \hgraph where every \hedge has exactly one source and every node has at most one outbound \hedge.
For the purpose of mapping, weights are per-\hedge and represent its spike frequency, how many spikes it is expected to carry in any one-second window \cite{AxonFlow}.

Formally, let $G = (N, E, \omega)$ be a \hgraph, with $N$ its set of nodes and $E$ its \hedges.
Each \hedge $e \in E$ connects a subset of nodes (pins) $e \subseteq N$ and has source $src(e)$, with $src : E \rightarrow N$ and destinations $dst(e)$, with $dst : E \rightarrow \mathcal{P}(N)$.
Where $\mathcal{P}(\cdot)$ is the power set and naturally $dst(e) = e - src(e)$.
Let $\omega : E \rightarrow \mathbb{R}$ assign weights -- spike frequencies -- to \hedges.
In addition, we denote a node $n$'s inbound \hedges as $in(n) = \{e \in E \sothat n \in dst(e)\}$ and its outbound \hedge as the singleton $out(n) = \{e \in E \sothat n = src(e)\}$, with $\mathcal{I}(n) = in(n) \cup out(n)$ as the set of \hedges incident to $n$.
Lastly, we define the neighbors of $n$ as $\mathcal{N}(n) = \{m \in e \sothat e \in \mathcal{I}(n)\} \setminus \{n\}$.

The model for NMH is a 2D lattice where each point represents a core: $H = \{(x, y) \sothat x \in \{1, \dots \text{width}\}, y \in \{1, \dots \text{height}\}\}$.
For a core $h \in H$, let $\mathcal{J}(h) = \{(h_x + 1, h_y), (h_x - 1, h_y), (h_x, h_y + 1), (h_x, h_y - 1)\} \cap H$ be the set of cores adjacent and connected to it.
Moreover, each core has a maximum number of neurons it can handle $\Omega$, axons it can receive spikes from $\Delta$, and synapses it can store $\Phi$.


\section{The Mapping Problem}\label{sec:mapping_problem}

\begin{figure}[t]
    \centering
    \vspace{-0pt} 
    \includegraphics[width=1.0\columnwidth]{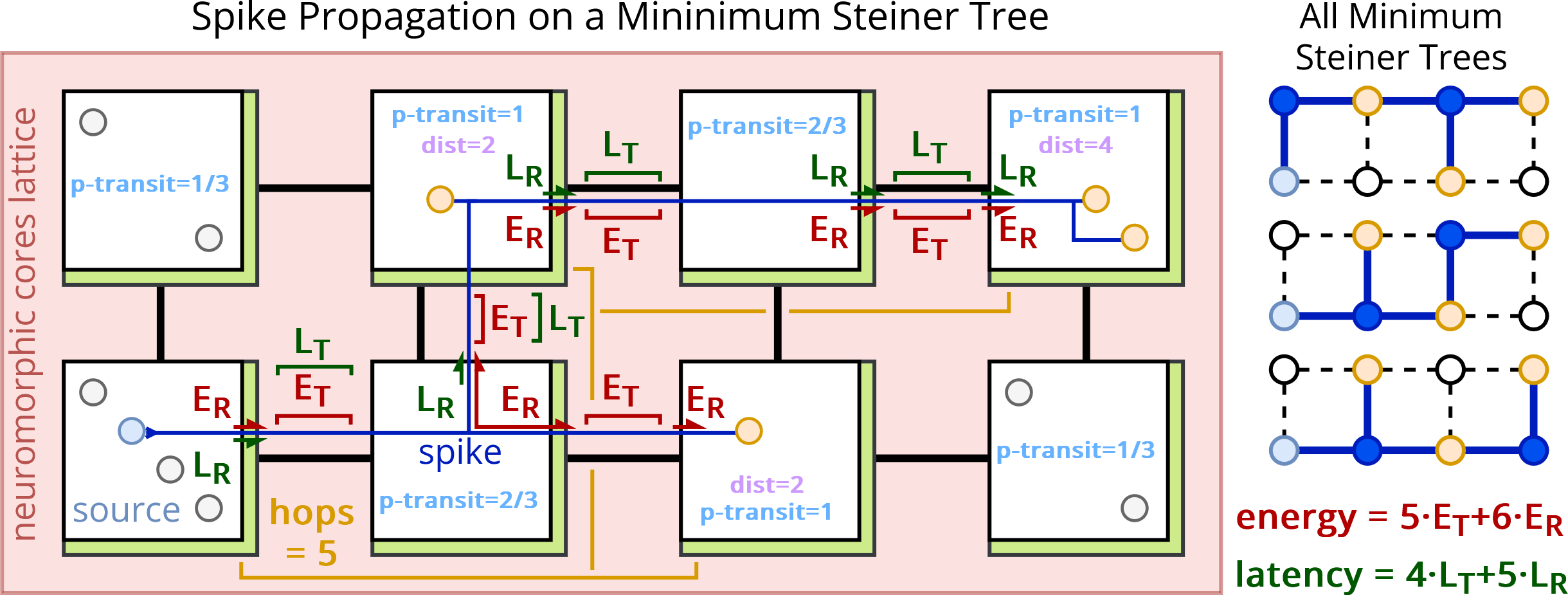}
    \vspace{-18pt}
    \caption{Cost model for a spike's propagation on NMH.}
    \vspace{-3pt}
    \label{fig:spike_propagation}
\end{figure}

The mapping problem consists of assigning each SNN neuron to a NMH core while minimizing the resulting spike traffic and can be addressed in two sub-problems.
First, partitioning groups neurons within core constraints while minimizing the volume of spikes crossing between groups.
Then, placement uniquely assigns groups to cores, minimizing the distance between densely communicating groups.
Our formulation is a complete rethinking for \hgraphs of the ideas in \cite{MappingVeryLargeSNNtoNHW, AxonFlow}.
Fig.~\ref{fig:mapping_overview} shows the full process.


\subsubsection*{Partitioning Definition}\label{subsec:partitioning_definition}

A partitioning of a SNN's \hgraph $G$ is a set $P \subset \mathcal{P}(N)$ of disjoint non-empty subsets of its nodes, such that $\bigcup_{p \in P} p = N$.
Equivalently, it can be defined by $\rho : N \rightarrow P$ where $\rho(n) = p$ iff $n \in p$.
Its objective is the connectivity, minimizing the weight of connections that are cut between partitions:
\begin{equation}\label{eq:connectivity}
    Conn_G(\rho) = \sum_{e \in E} \omega(e) \cdot \left( \abs{\{\rho(n) \sothat n \in e\}} - 1\right) \text{.}
\end{equation}
Hardware imposes three constraints per partition: a maximum number of nodes $\forall p \in P, \: \abs{p} \leq \Omega$; of distinct inbound \hedges, $\abs{\bigcup_{n \in p} in(n)} \leq \Delta$; and of inbound pins, $\sum_{n \in p} \abs{in(n)} \leq \Phi$.

By applying the $\rho$ map to both nodes and \hedges of $G$, we define another \hgraph among partitions: $G_P = \{P, A, \eta\}$.
Where $A = \{\rho(e) \sothat e \in E\}$ and $\eta : A \rightarrow \mathbb{R}$ is derived for every $a \!\in\! A$ as $\eta(a) = \sum_{e \in E \text{ s.t. } \rho(e) = a} \omega(e)$.
After breaking self-cycles by discarding destinations, $\:\!src(\cdot)$ and $dst(\cdot)$ analogously exist over $A$.

\subsubsection*{Placement Definition}\label{subsec:placement_definition}

A placement of the partitioned SNN \hgraph $G_P$ on a hardware lattice $H$ consists of an injective map $\gamma : P \rightarrow H$.
Local costs on every \hedge $a \in A$ or core $h \in H$ are threefold:
\begin{align}
    & \text{energy}(a) \;= hops(\gamma(a)) \cdot (E_R + E_T) + E_R \text{ ,} \label{eq:hedge_energy} \\
    & \text{latency}(a) \;= \!\! \max_{d \in dst(a)} \!\! dist(\gamma(src(a)), \gamma(d)) \cdot (L_R + L_T) + L_R \text{ ,} \label{eq:hedge_latency} \\
    & \text{congestion}(h) \;= \textstyle\sum_{a \in A} \: \eta(a) \cdot p\text{-}transit(h, \gamma(a)) \label{eq:core_congestion} \text{ .}
\end{align}
Where $dist : H \times H \rightarrow \mathbb{N}$ is the shortest path length on $H$ between two placed nodes; on a 2D lattice, $dist$ is the Manhattan distance.
Then, $hops : \mathcal{P}(H) \rightarrow \mathbb{N}$ counts the core-core links traversed by a spike to reach all destinations of an \hedge from its source.
And $p\text{-}transit : H \times \mathcal{P}(H) \rightarrow [0, 1]$ is the probability of a spike traversing core $h$ while it propagates between an \hedge's terminals.

By aggregating local costs, we define a global model of mapping performance.
Its quality metrics -- to minimize -- are defined as:
\begin{align}
    & \text{Tot. Energy} \;=\, \textstyle\sum_{a \in A} \: \eta(a) \cdot \text{energy}(a) \text{ ,} \\
    & \text{Avg. Latency} \;=\, \left( \nicefrac{1}{\sum_{a \in A}\eta(a)} \right) \cdot \textstyle\sum_{a \in A} \eta(a) \cdot \text{latency}(a) \text{ ,} \\
    & \text{Max. Congestion} \;=\, \textstyle\max_{h \in H} \: \text{congestion}(h) \text{ .}
\end{align}
These are analytical communication metrics, abstracting away platform-specific execution details.
Total energy captures only spike traffic, as neuron-update energy is unaffected by mapping.
Average latency is the frequency-weighted spike delivery completion time.
Maximum congestion tracks peak routing pressure across cores.

\begin{table}[b]
    \vspace{1pt} 
    \centering
    \begin{minipage}[t]{0.39\columnwidth}
        \centering
        \resizebox{\linewidth}{!}{
            \begin{tblr}{colspec={|X[0.5, c, m]|X[0.5, c, m]l|}, column{3} = {wd = 0pt, leftsep = 0pt, rightsep = 0pt}, rowsep = 0pt, width=1.25\linewidth} 
                \hline
                \SetCell[r=2]{c} \textbf{NMH~Cost} & \SetCell[r=2]{c} \textbf{Value} & \phantom{\textbf{Value}} \\
                \cline{2-3}
                & & \phantom{small} \\
                \hline
                $E_R$ & 1.7 pJ & \phantom{$width$,~$height$} \\
                \hline
                $L_R$ & 2.1 ns & \phantom{$\Omega$} \\
                \hline
                $E_T$ & 3.5 pJ & \phantom{$\Delta$} \\
                \hline
                $L_T$ & 5.3 ns & \phantom{$\Phi$} \\
                \hline
            \end{tblr}
        }
    \end{minipage}%
    \hspace{0.01\columnwidth}
    \begin{minipage}[t]{0.59\columnwidth}
        \centering
        \resizebox{\linewidth}{!}{
            \begin{tblr}{colspec={|X[c, m]|X[0.5, c, m]|X[0.5, c, m]|}, rowsep = 0pt, width=1.25\linewidth}
                \hline
                \SetCell[r=2]{c} \textbf{Constraint} & \SetCell[c=2]{c} \textbf{Value} & \\
                \cline{1-3}
                & small & large \\
                \hline
                \!\!\!\!\!\! \hfill \textsmaller[1]{neurons / core} \hfill $\Omega$ & 1024 & 4096 \\
                \hline
                \!\!\!\!\!\! \hfill \textsmaller[1]{axons / core} \hfill $\Delta$ & 4096 & 65536 \\
                \hline
                \!\!\!\!\!\! \hfill \textsmaller[1]{synapses / core} \hfill $\Phi$ & 16384 & 262144 \\
                \hline
                \textsmaller[1]{width~$\times$~height} & \SetCell[c=2]{c} 64$\:\times\:$64 & \\
                \hline
            \end{tblr}
        }
    \end{minipage}
    \vspace{-4pt} 
    \caption{Reference hardware costs \cite{Loihi} and constraints \cite{MappingVeryLargeSNNtoNHW}.}
    \vspace{-22pt} 
    \label{tab:hw_costs}
\end{table}

The definition of $hops$ and $p\text{-}transit$ depends on routing policies.
We therefore model multicast under a router-agnostic lower bound: each spike spans its endpoints with a minimum Steiner tree over the hardware mesh \cite{ForestAndSteinerTrees}.
Concrete routers may incur higher costs, but a lower Steiner cost means that the mapping leaves less irreducible multicast traffic for the implementation to realize.
Formally:
\begin{align}
    & T_H(hs) = \{ t \subseteq H \sothat hs \subseteq t \text{ and } t \text{ spans a tree in } H \} \text{ ,} \\
    & T_H^\star(hs) = \textstyle\argmin_{t \in T_H(hs)} \abs{t} \text{ ,} \label{eq:steiner_trees} \\
    & hops(hs) = \abs{t^\star} - 1 \text{\; for any \;} t^\star \!\in T_H^\star(hs) \text{ ,} \label{eq:hops} \\
    & p\text{-}transit(h, hs) = \abs{\{t \in T_H^\star(hs) \sothat h \in t\}} \,/\, \abs{T_H^\star(hs)} \text{ .} \label{eq:transit}
\end{align}
Where $T_H(hs)$ is the set of core sets that support a tree spanning terminals $hs$ over $H$.
Thus, any $t^\star \!\in T_H^\star(hs)$ encodes a minimum-size Steiner tree containing $\abs{t^\star}-1$ links.
Terminal roles as sources or destinations do not alter the $T_H^\star(hs)$ set.
Trees are represented by their core sets; edge variations that induce identical core traversals are disregarded, as congestion is modeled per-core.
See Fig.~\ref{fig:spike_propagation}.

Since Steiner tree construction is also NP-hard, we use it only to compute exact final metrics, after mapping.
Heuristics, instead, rely on proxy metrics for the number of hops and transit probability.

The hardware-specific costs $E_R$, $E_T$ and $L_R$, $L_T$ represent, respectively, the energy and latency for routing and transmitting a spike towards a core.
The final $E_R$ and $L_R$ terms in Eqs.~\ref{eq:hedge_energy} and \ref{eq:hedge_latency} account for the source core's dispatch cost.
Reference values for this work are based on the Loihi platform \cite{Loihi} and given in Tab.~\ref{tab:hw_costs}.

\subsubsection*{Capturing Hardware Behavior}\label{subsec:capturing_hardware_behavior}



During partitioning, \hedges natively expose intra-core spike replication.
When an \hedge reaches multiple destinations inside the same partition, connectivity still counts it as one cut, matching the cost of a single spike being delivered.

For placement, \hedges materialize as spike multicast trees.
Defining hops over all of an \hedge's pins does not just pull destinations closer to the source, but encourages the whole \hedge to contract, effectively allowing spikes to fork closer to their endpoints.

Consequently, minimizing connectivity (Eq.~\ref{eq:connectivity}) amounts to maximizing intra-core replication, while lower placement costs (Eqs.~\ref{eq:hedge_energy}, \ref{eq:hedge_latency}, \ref{eq:core_congestion}) reflect efficient and localized inter-core multicast.
Existing SNN mapping tools adopt a graph SNN model, thereby underexploiting both such dynamics \cite{AxonFlow}.
These nuances also distinguish SNN mapping from standard partitioning and placement formulations, and motivate algorithms specialized for hypergraph-defined objectives.

\section{GPU-based Mapping Pipeline}\label{sec:massive_parallelism}

The GPU execution model exposes hierarchical parallelism: in CUDA terminology, threads are organized into blocks, each internally scheduled as 32-thread warps under SIMT execution.

For an \hgraph $G(N,E,\omega)$, problem size scales with nodes $\abs{N}$, \hedges $\abs{E}$, and pins $\sum_{e\in E}\abs{e}$, whereas local structural parameters, namely \hedge cardinality $\abs{e}$ and node incidence degree $\abs{\mathcal{I}(n)}$, remain comparatively small in practice.
This separation makes the nested traversals used by \hgraph algorithms amenable to GPUs.
Outer traversals over $N$ or $E$ expose coarse-grained parallelism and map to blocks and warps; intermediate nested levels run as warp-synchronous serial loops; the innermost bounded expansion over pins or incidence lists is processed cooperatively within warps.
Thus, sequential work is confined to small neighborhoods, while most parallelism is extracted from the \hgraph's outer dimensions.



To support this traversal pattern, we store \hgraphs in compressed sparse form by concatenating local structures \cite{SIMDefficientGraphsOnGPU}.
In the resulting layout, each warp processes contiguous sub-arrays, ensuring coalesced accesses while limiting control divergence.

\subsection{Partitioning}\label{sec:partitioning}

\begin{figure*}[th]
    \centering
    \vspace{-2pt}
    \includegraphics[width=1.0\textwidth]{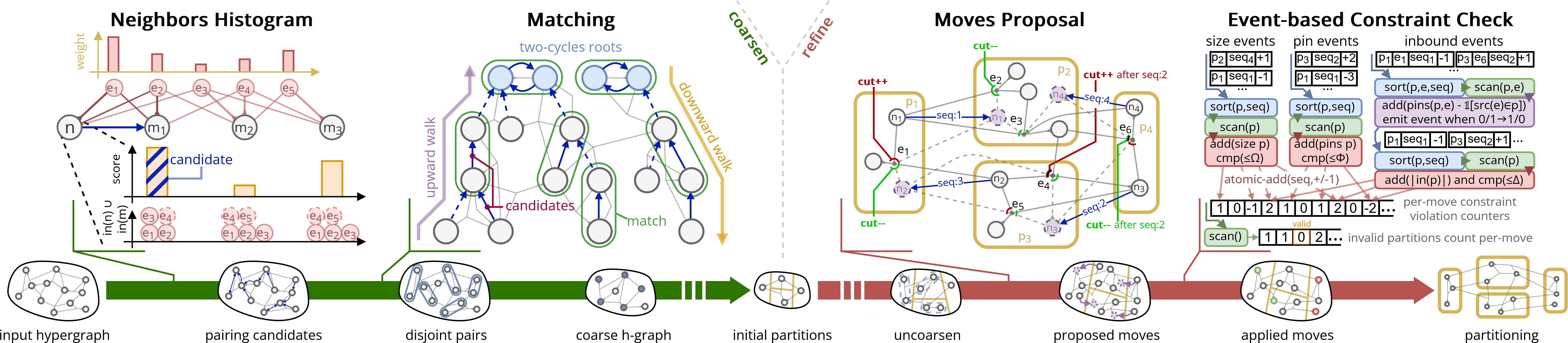}
    \vspace{-20pt} 
    \caption{Partitioning phases -- coarsening in green, uncoarsening and refinement in red -- and steps with key algorithm highlights.}
    \vspace{-14pt}
    \label{fig:partitioning_flow}
\end{figure*}

For partitioning, we adopt a multi-level approach, which is the leading technique to reach near-optimal results at scale \cite{AdvancesInHypergraphPartitioning, hMETIS_vlsi, KaHyPar, FMpartitioning}.
Ours is developed from \cite{AxonCUDA-IPDPS}, adding support for all constraints $\Omega, \Delta, \Phi$ via neighbor-counting and a simplified event-based pipeline.

The multi-level framework has two phases, as Fig.~\ref{fig:partitioning_flow} shows.
Through a series of coarsening levels, strongly connected nodes are merged into super-nodes, that form the next level's \hgraph.
When constraints forbid nodes from further merging, final super-nodes define initial partitions.
Then, as levels uncoarsen one by one, nodes are moved between partitions to reduce cut \hedges.

Each level's \hgraph is materialized from its predecessor while propagating each super-node's constrained state: size, inbound \hedges set, and pin count.
This way, all constraints are enforced on every super-node, guaranteeing partition validity throughout.

\subsubsection*{Coarsening}\label{subsec:coarsening}

Coarsening's goal is to form disjoint pairs of nodes of maximum total connection strength within constraints.
Each node $n \in N$ computes a histogram over its neighbors $m \in \mathcal{N}(n)$, accumulating their shared \hedge weight and shared inbound count $\abs{in(n) \cap in(m)}$.
This overlap term lets us test the merged inbound set against $\Delta$.
Neighbors violating node, pin, or inbound \hedge limits are discarded, and $n$ selects the valid maximum as its pairing candidate, tie-breaking by node id:
\begin{equation}\label{eq:neigh_histogram}
    \begin{aligned}
        & cand(n) = \textstyle\argmax^{\text{valid}}_{m \in \mathcal{N}(n)} score(n, m) \\
        & \text{where } score(n, m) = \textstyle\sum\{\omega(e) \sothat e \in \mathcal{I}(n),\, m \in e\} \text{ .}
    \end{aligned}
\end{equation}

To form super-nodes, candidate pairs must be disjoint; therefore, coarsening requires a high-weight matching over the directed candidate graph.
Since each node has at most one candidate, this graph is functional.
Moreover, histograms are symmetric and candidates are maxima, so selected scores are non-decreasing along arcs, and tie-breaking restricts cycles to two nodes.
Thus, the graph is a pseudo-forest rooted at two-cycles.
We therefore use the simultaneous-walk matching procedure of \cite{AxonCUDA-IPDPS}.
A thread per node walks upward along candidate arcs to its root two-cycle, atomically acquiring candidates with conflicts resolved by score.
After synchronization, two-cycles are matched and threads walk downward by retracing their paths, matching nodes that still hold their acquisitions when revisited.
The resulting matching is characterized recursively as:
\begin{equation}\label{eq:matching}
    \hspace{-0.4em} match(n) = \begin{cases}
        cand(n) \hspace{9.25em} & \hspace{-9.25em} \begin{aligned}
            & \text{if } cand(cand(n)) = n \\[-4pt]
            & \text{or } match(cand(n)) = n
        \end{aligned} \text{\;,} \\[4pt]
        \textstyle\argmax_{m \text{ s.t. } cand(m) = n} score(n, m) & \hspace{-0.3em} \text{if }\!\!\: m \!\!\:\text{ exists\,.\!}
    \end{cases}
\end{equation}

Lastly, the coarse \hgraph is built, or, if no candidates are proposed, an initial $\rho$ is formed with a partition per super-node.

\subsubsection*{Uncoarsening and Refinement}\label{subsec:uncoarsening_and_refinement}

With a $\rho$ providing provisional partitions, super-nodes come undone level by level while refinement moves nodes so to disconnect \hedges from as many partitions as possible.
For a disconnection to happen, a node must be the last one for an \hedge in a partition.
Thus, to find favorable moves, the pin count per partition is precomputed as $pins(p, e) = \abs{\{n \in e \sothat n \in p\}}$.

Each node $n$ computes the spared cuts cost for leaving its current partition as $save(n) \!=\! \sum\{\omega(e) \sothat e \!\in\! \mathcal{I}(n),\, pins(\rho(n), e) \!=\! 1\}$.
At the same time, for all partitions $p$ besides its own, it finds the cost for entering them as $loss(n, p) \!=\! \sum\{\omega(e) \sothat e \!\in\! \mathcal{I}(n),\, pins(p, e) \!=\! 0\}$.
Now $save(n) - loss(n, p)$ is the connectivity gain for moving to $p$; each $n$ proposes the best such move that is by itself feasible.

While every node proposes a move, not all can be applied at once, as mutual interference could break their favorability and validity.
We therefore greedily sort moves by decreasing gain \cite{HyperG} and interpret refinement as selecting the best improving prefix of this sequence.
To evaluate each prefix, we first recompute every move's gain assuming all preceding moves have already been applied.
A prefix sum of these updated gains identifies the cumulative improvement of each prefix.
What remains is to determine, in the same move order, which prefixes produce a valid state.

Validity is determined by emitting sparse events along the move sequence.
Each move emits signed events for changes in the constrained quantities directly affected by its node: partition size and total inbound pin count.
After grouping events by partition and scanning them in move order, crossings of the $\Omega$ and $\Phi$ thresholds update a per-move violation counter.
Events for the inbound set size require one more step, as they arise only when an \hedge connects or disconnects from a partition.
For each affected pair $(p, e)$, moves emit signed events describing the variation of $pins(p, e)$.
Their prefix sum yields the running value of $pins(p,e)$, from which we test whether $e$ is inbound to $p$ as $pins(p, e) - \mathbbm{1}[src(e) \in p] > 0$.
Transitions of this condition generate signed events for the inbound set size of $p$.
Scanning these events by partition detects crossings of $\Delta$ and updates the same per-move violation counters.

A move is valid iff the prefix sum of violation counters is zero at its position.
We enact moves up to the highest cumulative gain prefix that is both improving and valid.


\subsection{Placement}\label{sec:placement}

\begin{figure*}[th]
    \centering
    \vspace{-2pt}
    \includegraphics[width=1.0\textwidth]{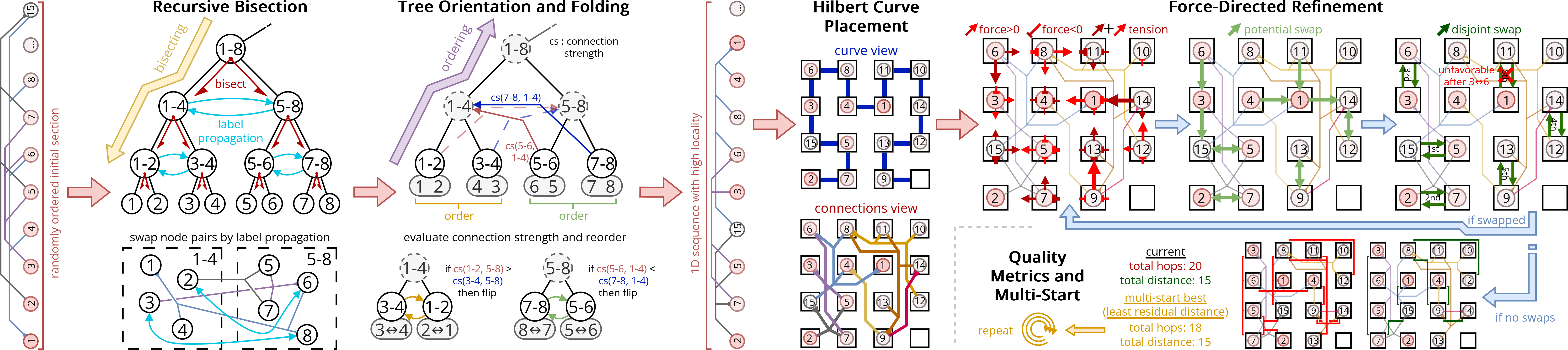}
    \vspace{-18pt}
    \caption{Placement phases -- initial 1D placement on the left, Hilbert curve and refinement on the right -- and steps with node swap highlights.}
    \vspace{-14pt}
    \label{fig:placement_flow}
\end{figure*}

For placement, we adapt several ideas from \hgraph combinatorial optimization \cite{KaHyPar_origins, MtKaHypar-Steiner, ForestAndSteinerTrees, PairExchangeAssignment} and prior SNN mapping tools \cite{MappingVeryLargeSNNtoNHW, MappingVeryLargeSNNtoNHWv2}.

Placement is performed in two phases: the construction of an initial solution, followed by its iterative refinement, see Fig.~\ref{fig:placement_flow}.
The initial solution is obtained greedily, by flattening $H$ to a single dimension and constructing a high-locality sequence of nodes over it.
Subsequently, refinement lets \hedges "pull" on their nodes, inducing forces that promote swaps between adjacent cores.

Reasonably assuming all partitions to be near-full, forming $G$ into $G_P$ shrinks the node count by a factor of about $\min(\Omega, \nicefrac{\Delta}{\max\abs{in(n)}})$.
With $G_P$ relatively small, a multi-start method becomes practical: we refine $\Pi$ placements concurrently from different initial conditions, and ultimately select the best one.
For this work we set~$\Pi \!=\! 64$.

\subsubsection*{Initial Placement}\label{subsec:initial_placement}

First, $H$ is projected into a 1D sequence via the Hilbert curve, which preserves most of its 2D locality \cite{MappingVeryLargeSNNtoNHW}.
Then, a sequence of nodes minimizing the distance between strongly connected neighbors is built and nodes are assigned over $H$ in order.
This is a 1D reduction of the problem, that we solve via recursive bisection followed by a tree-based orientation optimization.

We distribute nodes over internally ordered sections $\{s_0, s_1, \dots\} \!\subseteq\! \mathcal{P}(P)$.
Initially, all nodes belong to the same section $s_0$, in random order.
At each recursion, every section $s_i$ is split halfway in two child sections $s'_{2i}$ and $s'_{2i+1}$ s.t. $s_i = s'_{2i} \!\! \concat s'_{2i+1}$.
Follow several rounds of label propagation \cite{AdvancesInHypergraphPartitioning} that swap nodes between pairs of child sections, minimizing the total weight of \hedges cut in the bisection.

The recursion halts once each section contains exactly one node, leaving behind a binary tree where leaves coincide with single nodes and branches with bisected sections.
Now the sequence $s_0, s_1, \dots, s_{\abs{P}-1}$ of leaves already exhibits some locality; however, the order of child sections on each branch is still arbitrary.

To construct the final node order from sections, the above bisection tree is ascended while folding its branches.
Starting from leaves, at each step, pairs of sibling sections are concatenated, and their relative order is chosen by which one has the stronger total connection to the nodes in their parent's sibling section.
Intuitively, sections are oriented so that their most strongly connected sides face each other as folding reaches higher levels of the tree.
Moreover, whenever two sections flip positions their internal node order is also reversed.
This effectively traps the strong connections identified at previous tree levels within each compound section.
By the end, the root section contains a sequence with high local connectivity.

We simultaneously construct $\Pi$ initial placements in this way.

\subsubsection*{Swap Proposals}\label{subsec:swaps_proposal}

Each node $p \in P$, currently placed at $h = \gamma(p)$, computes a force towards each adjacent core.
A $force : P \times H \rightarrow \mathbb{R}$ represents the total weight of hops that could be spared in reaching all of $p$'s neighbors by moving $p$ to any adjacent place:
\begin{equation}\label{eq:force}
    \begin{gathered}
        \forall k \in \mathcal{J}(h), \; force(p, k) = \\
        = \sum_{a \in \mathcal{I}(p)} \!\! \eta(a) \, \cdot \!\!\!\! \sum_{q \in a \setminus \{p\}} \!\! \big(\, dist(h, \gamma(q)) - \max(dist(k, \gamma(q)), 1) \,\big)
    \end{gathered}
\end{equation}
being zero for any other $k$.
A strictly positive force marks $k$ as a desirable placement for $p$.

By summing over all node-neighbor pairs, forces jointly capture sources pulling on destinations and destinations pulling on each other.
Src-dst terms directly promote lower distance connections, addressing the latency objective in Eq.~\ref{eq:hedge_latency}.
Dst-dst terms instead encourage the locality of pins within each \hedge, fostering multicast and acting as a surrogate for reducing $hops$ in accord with Eq.~\ref{eq:hops}.


With the focus shifting on cores, let us define the inverse placement $\pi : H \rightarrow P$ as $\pi(h) = p \text{ if } \gamma(p)=h;\; \bot \text{ if } h \text{ is empty}$.

The lattice is densely packed with nodes, therefore a node changing place means swapping it and the one currently occupying such place, if any.
Hence, any force in favor of a new placement will be met by the opposing force of the node that is already there.
The resulting $tension : H \times H \rightarrow \mathbb{R}$ between adjacent places is:
\begin{equation}\label{eq:tension}
    tension(h,k) = \!\!\mathop{\mathbbm{1}}\limits_{\pi(h)\neq \bot}\!\!\!\!\!\! \cdot \,\, force(\pi(h), k) + \!\!\mathop{\mathbbm{1}}\limits_{\pi(k)\neq \bot}\!\!\!\!\!\! \cdot \,\, force(\pi(k), h) \text{ .}
\end{equation}
Naturally, the tension is symmetric, $tension(h, k) = tension(k, h)$.
A positive tension signifies that a swap would overall decrease the distance covered to connect the affected nodes.
Consequently, the tension is a compound local proxy of both energy and latency based on \hedges locality \cite{AxonFlow}.
Thus, every node $p$ proposes a swap with $\argmax_{k \in H} tension(\gamma(p), k)$, tie-breaking lexicographically over $k$.

Forces are computed with a single neighbors iteration: distances from the node's placement to each pin are computed once, then reused to derive distances from adjacent placements incrementally.
A second pass over adjacent core pairs updates forces to tensions.

\subsubsection*{Placement Refinement}\label{subsec:placement_refinement}

For several swaps to be performed together, their cores must be disjoint.
Hence, a high-total-tension matching is required between cores over the graph induced by proposed swaps.
The same conditions as in Sec.~\ref{subsec:coarsening} apply, with non-decreasing tensions along paths of proposed swaps leading to a two-cycle pseudo-forest.
Consequently, we adopt the same matching procedure.

Even with disjoint swaps, applying all of them at once may not be favorable, as they might interfere with each other's tension.
Thus, we sort swaps by tension into a sequence, following the same idea as in Sec.~\ref{subsec:uncoarsening_and_refinement}.
Then, another neighborhood traversal updates each tension as if all swaps before it were already performed.
Subsequently, a prefix sum of tensions and a reduce-max operation yield the best improving prefix in the sequence.
All selected swaps are applied at once, with an update to $\gamma$ and $\pi$.

Refinement repeats until no swaps occur.
Convergence is guaranteed by the pairwise-distance descent argument of \cite{MappingVeryLargeSNNtoNHW}.
The best multi-start result is chosen by minimum total residual distance.


\section{Experimental Evaluation}\label{sec:experimental}

\begin{figure}[b]
    \centering
    \vspace{3pt} 
    \includegraphics[width=\columnwidth]{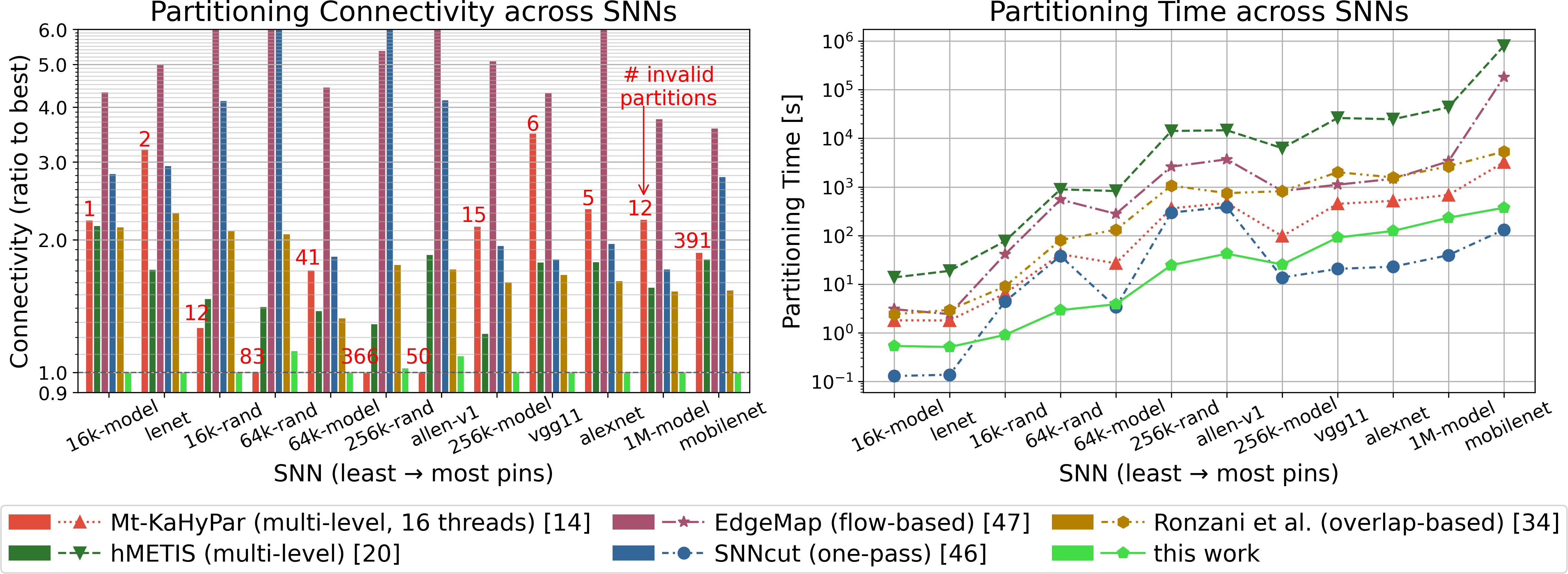}
    \vspace{-22.5pt} 
    \caption{Partitioning results comparison. Lower is better.}
    \vspace{-6pt} 
    \label{fig:part_results}
\end{figure}

\begin{figure*}[t]
    \centering
    \begin{minipage}[t]{0.805\textwidth}
        \centering
        \resizebox{\linewidth}{!}{
            \begin{tblr}{colspec={|X[0.3,c,m]|X[1.68,l,m]|*{12}{X[0.85,c,m]}|}, row{1} = {c}, width=1.4\linewidth, rowsep = 0.4pt}
                \hline
                \SetCell[c=2]{c}{\textbf{Network}} &
                & \clap{16\kk-model} & \clap{lenet} & \clap{16\kk-rand} & \clap{64\kk-rand} & \clap{64\kk-model} & \clap{256\kk-rand} & \clap{allen-v1} & \clap{256\kk-model} & \clap{vgg11} & \clap{alexnet} & \clap{\!\!\:1\MM-model} & \clap{mobilenet} \\
                \hline
                \SetCell[c=2]{l}\textbf{Topology} &
                & acyclic & acyclic & cyclic & cyclic & acyclic & cyclic & cyclic & acyclic & acyclic & acyclic & acyclic & acyclic \\
                \SetCell[c=2]{l}\textbf{Target constraints} &
                & small & small & small & small & small & small & large & large & large & large & large & large \\
                \hline
                \SetCell[r=3]{c}{\hspace*{-5pt}\rotatebox{90}{\!\!\shortstack{\clap{\textsmaller[1]{original}}\\($G$)}}}
                & \textbf{\!Node count}
                & 20\kk & 14\kk & $2^{14}$ & $2^{16}$ & 110\kk & $2^{18}$ & 231\kk & 216\kk & 194\kk & 208\kk & 302\kk & 6.9\MM \\
                & \textbf{\!Pin count}
                & 766\kk & 875\kk & 2.1\MM & 12.6\MM & 23\MM & 67.4\MM & 70\MM & 90\MM & 133\MM & 145\MM & 256\MM & 577\MM \\
                & \textbf{\!Mean \!\!\:node \!\!\:deg.\!\!}
                & 37.3 & 63.2 & 128 & 192 & 210.3 & 256 & 304.7 & 417.2 & 688.3 & 696.2 & 848.1 & 83.5 \\
                \hline
                \SetCell[r=3]{c}{\hspace*{-5pt}\rotatebox{90}{\!\!\shortstack{\clap{\textsmaller[1]{partitioned}}\\($G_P$)}}}
                & \textbf{\!Node count}
                & 23 & 16 & 19 & 120 & 121 & 653 & 60 & 60 & 57 & 59 & 84 & 1.7\kk \\
                & \textbf{\!Pin count}
                & 535 & 298 & 19.2\kk & 412\kk & 5.7\kk & 2.3\MM & 558\kk & 2.4\kk & 1.3\kk & 2.1\kk & 5.9\kk & 5.4\MM \\
                & \textbf{\!Mean \!\!\:node \!\!\:deg.\!\!}
                & 2.63 & 2.56 & 4.12 & 6.32 & 2.88 & 8.85 & 10.6 & 2.64 & 2.76 & 2.73 & 4.27 & 10.0 \\
                \hline
            \end{tblr}
        }
        \vspace{-2pt}
        \captionof{table}{SNN hypergraphs, original and partitioned (with lowest-connectivity available), used in the experiments.}
        \label{tab:snns}
    \end{minipage}
    \hfill
    \begin{minipage}[t]{0.18\textwidth}
        \centering
        \vspace{-45pt}
        \includegraphics[width=\linewidth]{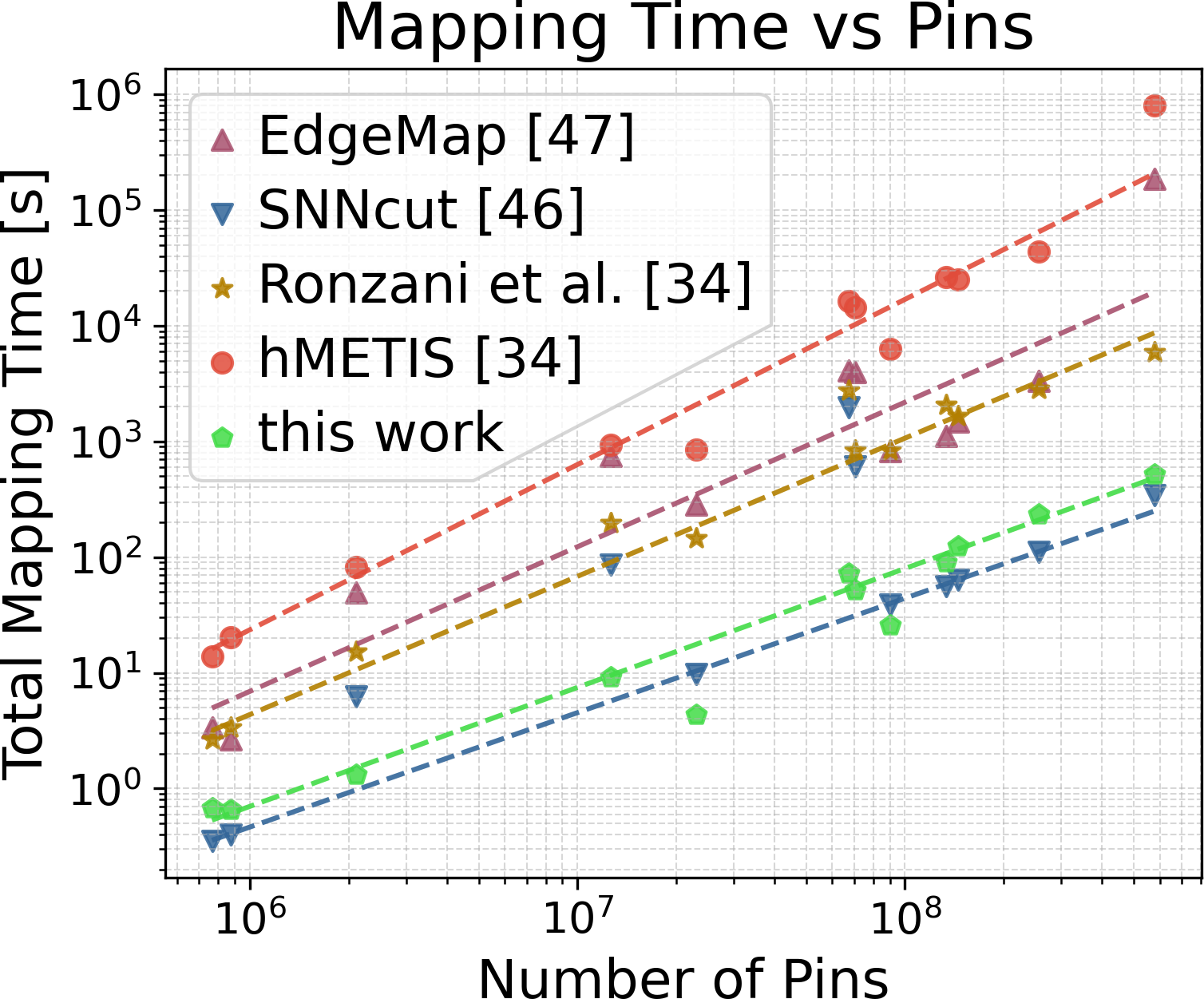}
        \vspace{-22.5pt}
        \captionof{figure}{\setlength{\baselineskip}{5pt}Scalability of mapping tools over pins.}
        \label{fig:scaling}
    \end{minipage}
    \vspace{-15pt}
\end{figure*}

\begin{figure*}[t]
    \centering
    \vspace{-6pt}
    \includegraphics[width=1.0\textwidth]{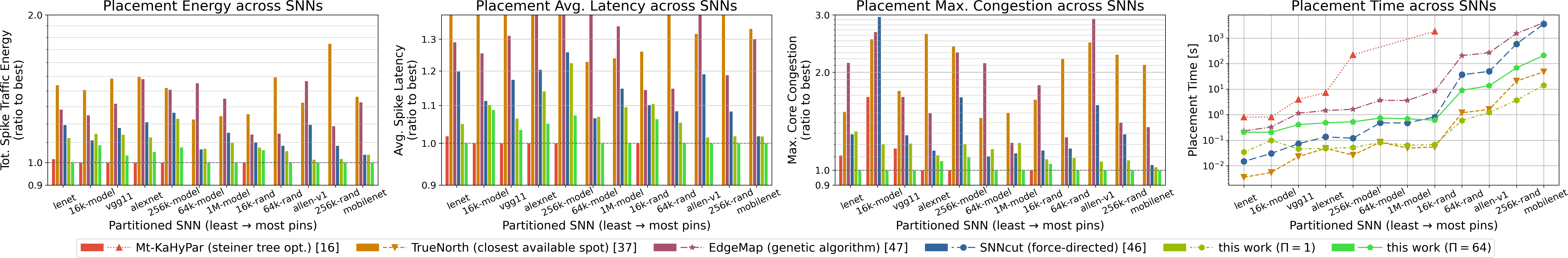}
    \vspace{-22.5pt} 
    \caption{Placement results comparison. Algorithms start from the same partitioning: the lowest-connectivity available one. Lower is better.}
    \label{fig:plac_results}
    \vspace{-3pt}
\end{figure*}

\begin{figure*}[t]
    \centering
    \vspace{-6pt}
    \includegraphics[width=1.0\textwidth]{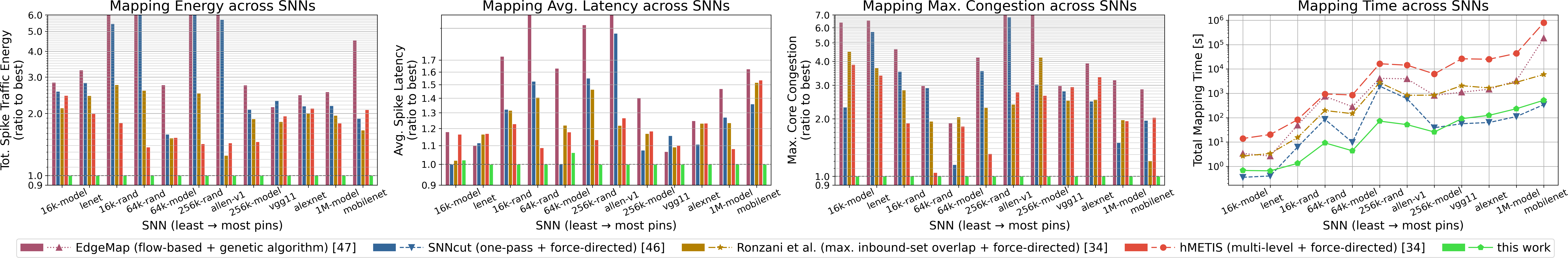}
    \vspace{-22.5pt} 
    \caption{Mapping -- partitioning and placement -- results comparison with four sequential methods across twelve SNNs. Lower is better.}
    \label{fig:map_results}
    \vspace{-14pt}
\end{figure*}

To evaluate our approach, we compare with existing mapping tools, all running sequentially on CPU.
SNNcut \cite{SNNcut} employs a one-pass node-order partitioner (extended to cyclic networks via greedy ordering \cite{AxonFlow}) and a Hilbert-curve force-directed placement scheme \cite{MappingVeryLargeSNNtoNHW, MappingVeryLargeSNNtoNHWv2}, on which ours is based.
Ronzani et al. \cite{AxonFlow} use greedy inbound set overlap maximization for partitioning and SNNcut's same placement algorithm.
EdgeMap \cite{EdgeMap} uses a streaming flow-based partitioner and a genetic placement algorithm.
For placement only, we also include TrueNorth's internal method \cite{TrueNorthEcosystem}.
DFSynthesizer \cite{DFSynthesizer} and others \cite{SpiNeMap, SNEAP} have been excluded because already dominated by the selected tools \cite{SNNcut, EdgeMap} or missing public artifacts.

We further compare with two multi-level \hgraph partitioning tools: the classic hMETIS \cite{hMETIS_k_way} sequential algorithm, adapted to our constraints \cite{AxonFlow}, and the multi-threaded Mt-KaHyPar \cite{MtKaHyPar}.
In addition, Mt-KaHyPar supports a mapping mode under the Steiner tree objective \cite{MtKaHypar-Steiner}, against which we compare for placement.
Since Mt-KaHyPar does not support the distinct inbound \hedges constraint, we annotate the number of invalid partitions it produces and use it as an optimistic reference on \hgraphs.
To our knowledge, no GPU baseline exists for our tasks and constraints, e.g. HyperG \cite{HyperG} and gHyPart \cite{gHyPart} only target $k$-way balanced partitioning.

Benchmarks comprise twelve publicly available SNNs \cite{BenchmarkSNNs}, reported in Tab.~\ref{tab:snns} and ranging from million-scale to 577\MM pins.
They cover both ANN-derived acyclic feedforward networks, including the VGG-11-scaled \texttt{-model} family, and cyclic small-world networks, namely the Allen~V1 \cite{AllenV1} and \texttt{-rand} networks patterned after it.

NMH configurations follow Tab.~\ref{tab:hw_costs}: networks use the "small" setting up to $2^{26}$ pins and the "large" one beyond it, thereby evaluating each network on a realistically capable system for its size \cite{NeuromorphicHardwareSurvey}.

Following prior SNN mapping studies \cite{AxonFlow, MappingVeryLargeSNNtoNHW, MappingVeryLargeSNNtoNHWv2, SNNcut}, we evaluate all methods under the analytical communication model of Sec.~\ref{sec:mapping_problem}, with hardware costs from Tab.~\ref{tab:hw_costs}.
Using the same setup for every baseline, the resulting metrics focus the comparison on mapping-dependent spike-routing costs rather than hardware measurements.

Experiments ran on an EPYC 7453 @ 2.75GHz CPU and an A100-SXM4-40GB GPU.
Timings are end-to-end wall-clock, exclude post-hoc metric extraction, and average 5 runs with negligible variance.

\subsubsection*{Results Discussion}\label{sec:results_discussion}

We perform three sets of experiments.
First, partitioning methods are evaluated based on their connectivity and runtime.
Second, placement techniques are fairly assessed by fixing their input to the same partitioning, the lowest-connectivity one available.
Third, we compare the different pipelines end-to-end.

Partitioning results are presented in Fig.~\ref{fig:part_results}.
Our approach consistently outperforms SNN mapping tools, that yield a $1.3$-$15.9\times$ higher connectivity.
Our results also improve on \hgraph partitioning tools: hMETIS's mean connectivity is $1.6\times$ higher and Mt-KaHyPar's is $1.9\times$ so.
Just the latter achieves better results on three occasions, but while producing tens of invalid partitions.

Looking at execution time, our GPU-parallel approach shows a $25$-$2\text{k}\times$ speedup over the sequential hMETIS, and $3$-$15\times$ over the multi-threaded Mt-KaHyPar (16 threads), both multi-level schemes too.
We are also $5\times$ or faster than other mapping tools that iterate over pins (EdgeMap, Ronzani et al.), and at most $6\times$ slower than the single iteration over nodes of SNNcut.
Notably, small-world networks slow down other methods over irregular traversals or node ordering, whereas our runtime is mostly unaffected by topology.

Placement-only results are reported in Fig.~\ref{fig:plac_results}.
With multi-start ($\Pi = 64$), our approach outperforms existing SNN mapping tools: their mappings incur $1.02$-$1.75\times$ the energy, $1.02$-$1.76\times$ the avg. latency, and $1.04$-$2.96\times$ the max. core congestion, with execution times bracketing ours ($0.04$-$22\times$).
Remarkably, even without multi-start ($\Pi = 1$), our method already improves over most baselines while being nearly the fastest, signaling the robustness of our initial placement strategy.
Increasing $\Pi$ beyond $64$ yields negligible gains while incurring a proportional runtime overhead due to limited GPU resources, and is therefore not considered further.

Mt-KaHyPar provides a useful reference for attainable placement quality, showing that mappings with metrics up to $10\%$ lower than ours are sometimes possible.
However, reaching them incurs prohibitively high execution times, and its placement algorithm is limited to at most 64 lattice nodes, hence the missing results.
As such, our method remains the most effective scalable solution.

All independent improvements transfer over to end-to-end mapping quality, reported in Fig.~\ref{fig:map_results}.
There, our pipeline demonstrates the best mappings overall.
Taking, for each SNN and metric, the best competing result normalized to ours, existing tools still incur a mean $1.72\times$ energy, $1.18\times$ avg. latency, and $1.98\times$ max. congestion.

Our total runtime continues to exhibit a $4$-$1.5\text{k}\times$ speedup over non-trivial methods, while being competitive with SNNcut's greedy heuristics.
As shown in Fig.~\ref{fig:scaling}, mapping time generally scales near-linearly with pin count; our parallel solution preserves this trend while shifting the curve downward.
In particular, on the largest 577\MM-pins \texttt{mobilenet} SNN, we retain an execution time of a few minutes and GPU memory usage peaks at 28GB, while halving all mapping costs compared to SoTA methods.

\section{Conclusion}\label{sec:conclusion}

In this work, we presented the first GPU-accelerated pipeline for mapping spiking neural networks on neuromorphic hardware.
It achieves the highest mapping quality among the evaluated SoTA baselines while greatly reducing time-to-solution.
Hence, the scalability of our approach positions it to handle ever-growing networks, with brain-scale systems as the long-term target.
Having established mapping quality within this analytical framework, evaluation on hardware platforms is being pursued, pending integration with their software stacks.
Our tools are available as open-source \cite{AxonCUDARepo}.

\bibliographystyle{ACM-Reference-Format}
\bibliography{biblio_short}

\end{document}